\documentclass[journal]{IEEEtran}

\usepackage{xcolor,soul,framed} 

\colorlet{shadecolor}{yellow}
\usepackage[pdftex]{graphicx}
\graphicspath{{../pdf/}{../jpeg/}}
\DeclareGraphicsExtensions{.pdf,.jpeg,.png}

\usepackage[cmex10]{amsmath}
\usepackage{array}
\usepackage{mdwmath}
\usepackage{mdwtab}
\usepackage{eqparbox}
\usepackage{url}
\usepackage{siunitx}
\usepackage{comment}
\usepackage{booktabs}
\usepackage{makecell}
\usepackage{stfloats}

\begin{document}
\bstctlcite{IEEEexample:BSTcontrol}
    \title{1D Thermoembolization Model Using CT Imaging Data for Porcine Liver}
  \author{Rohan Amare, Danielle Stolley, Steve Parrish, Megan Jacobsen, Rick Layman, Chimamanda Santos, Beatrice Riviere, Natalie Fowlkes, David Fuentes, Erik Cressman

  \thanks{Manuscript received \emph{DATE}. This work was supported in part by an Institutional Research Grant (IRG) and the Tumor Measurement Initiative through The University of Texas MD Anderson Cancer Center STrategic Initiative DEvelopment Program (STRIDE). It was also supported by the NIH under award numbers P30CA016672, R01CA201127, and R21CA260016, and the NSF under award numbers NSF-2111147 and NSF-2111459. CS is supported by the UPWARDS Training Program (Underrepresented Minorities Working Towards Research Diversity in Science; award number 1R25CA240137-01A1) and by a Cancer Research \& Prevention Institute of Texas Research Training Award (RP210028). }
  \thanks{\emph{Corresponding author: David Fuentes and Erik Cressman}}
  \thanks{Rohan Amare is with the Department of Imaging Physics, the University of Texas MD Anderson Cancer Center, Houston, TX, USA (e-mail: RPAmare@mdanderson.org).}
  \thanks{Danielle Stolley is with the University of Texas MD Anderson Cancer Center, Houston, TX, USA.}%
  \thanks{Steve Parrish is with the Department of Interventional Radiology, the University of Texas MD Anderson Cancer Center, Houston, TX, USA}
  \thanks{Megan Jacobsen is with the Department of Imaging Physics, the University of Texas MD Anderson Cancer Center, Houston, TX, USA.}
  \thanks{Rick Layman is with the University of Texas MD Anderson Cancer Center, Houston, TX, USA.}
  \thanks{Chimamanda Santos is with the University of Texas MD Anderson Cancer Center, Houston, TX, USA.}
  \thanks{Beatrice Riviere is with the Department of Computational Applied Mathematics \& Operational Research, Rice University, Houston, TX, USA.}
  \thanks{Natalie Fowlkes is with the Department of Veterinary Medicine and Surgery, the University of Texas MD Anderson Cancer Center, Houston, TX, USA.}
  \thanks{David Fuentes is with he Department of Imaging Physics, the University of Texas MD Anderson Cancer Center, Houston, TX, USA (e-mail: DTFuentes@mdanderson.org).}
  \thanks{Erik Cressman is with the Department of Interventional Radiology, the University of Texas MD Anderson Cancer Center, Houston, TX, USA (e-mail: ECressman@mdanderson.org)}

}

\markboth{IEEE TRANSACTIONS 
}{Amare \MakeLowercase{\textit{et al.}}: }

\maketitle

\begin{abstract}
\emph{Objective}: Innovative therapies such as thermoembolization are expected to play an important role in improving care for patients with diseases such as hepatocellular carcinoma. Thermoembolization is a minimally invasive strategy that combines thermal ablation and embolization in a single procedure. This approach exploits an exothermic chemical reaction that occurs when an acid chloride is delivered via an endovascular route. However, comprehension of the complexities of the biophysics of thermoembolization is challenging. Mathematical models can aid in understanding such complex processes and assisting clinicians in making informed decisions. In this study, we used a Hagen-Poiseuille 1D blood flow model to predict the mass  transport and possible embolization locations in a porcine hepatic artery. 
\emph{Method}: The 1D flow model was used on imaging data of \emph{in-vivo} embolization imaging data of three pigs. The hydrolysis time constant of acid chloride chemical reaction was optimized for each pig, and LOOCV method was used to test the model's predictive ability.
\emph{Conclusion}:This basic model provided a balanced accuracy rate of \SI{66.8}{\percent} for identifying the possible locations of damage in the hepatic artery. Use of the model provides an initial understanding of the vascular transport phenomena that are predicted to occur as a result of thermoembolization. 
\end{abstract}


%
\IEEEpeerreviewmaketitle


\section{Introduction}

\IEEEPARstart{H}{epatocellular} carcinoma (HCC) is a major global health issue \cite{jemal2011global, bray2024global}. The annual incidence is high, estimated at more than 850,000 cases globally, and unlike with many cancers, its incidence is increasing as the prevalence of risk factors such as nonalcoholic fatty liver disease rise. Treatments of HCC vary widely depending on the tumor stage and degree of underlying liver disease. Surgery (partial hepatectomy) is potentially curative under ideal conditions, in which the patient has a single mass under 5 cm in diameter confined to the liver with no evidence of invasion of nearby hepatic vasculature. Liver function may be preserved after surgery in these optimal cases. Unfortunately, this ideal surgical population represents only \SI{5}{\percent} of cases in the United States \cite{pollock2004uicc, ikai1998surgical, vauthey2002simplified}; HCC is more frequently diagnosed in later stages because it lacks characteristic and specific symptoms during its early stages \cite{kew1971diagnosis}. Ablation and embolization are the two most common minimally invasive methods used in treating unresectable HCC in appropriate patients according to numerous algorithms. These are established therapies with a known survival advantage \cite{nault2018percutaneous, llovet2002arterial}. In particular, thermally ablative therapies, including radiofrequency ablation, microwave ablation, cryotherapy, and laser ablation, use thermal energy to destroy the diseased tissue and a margin of surrounding tissue that contains microscopic disease. Unfortunately, incomplete ablation is more prevalent than commonly believed \cite{moussa2014radiofrequency, thompson2014heat}. 

Thermoembolization \cite{fuentes2020mathematical, cressman2018image, cressman2018feasibility, cressman2018first} , which was first reported in 2018, is a novel conceptual transarterial approach to cancer treatment in which a bolus of acid chloride dissolved in an inert oily solvent delivers a reagent, resulting in an exothermic chemical reaction. This approach is unique, as it combines the benefits of embolic as well as thermal and chemical ablative therapy modalities and offers several advantages over current techniques. Specifically, the target tissue and vascular bed are subjected to simultaneous hyperthermia, ischemia, and chemical denaturation in a single procedure. Intuitively, embolic effects of this technique reduce blood flow near the burn zone and thus can reduce major heat-sink limitations observed with conventional liver ablation techniques. Delivery of the acid chloride dissolved in bolus to the target tumor is achieved through selective catheterization of the feeding vessel. Furthermore, inflammation in the periphery of the burn zone can enhance delivery of chemical denaturant byproducts that may synergistically increase the diameter of the lethal zone of this therapy. A relatively high concentration of these reaction byproducts is left behind in a localized region of a devascularized treatment area and serves as a local diffusion reservoir of chemical denaturant that potentially could decrease the risk of local HCC recurrence, a common problem with thermal ablation and embolization techniques.

In thermoembolization, the thermoembolic bolus is delivered through a small catheter in the target artery. The oily solvent delays the exothermic chemical reaction through equilibrium with diffusion and allows time for the acid chloride to reach the target tissue. As it diffuses from the oily solvent, the acid chloride reacts vigorously with any water or available functional groups present in the tissue and simultaneously generates an acidic local environment. This exothermic hydrolysis of acid chloride offers a number of different avenues for local tissue destruction based on the distribution of the resulting heat and reaction byproducts. Prior efforts \cite{cressman2018feasibility, cressman2018first} have demonstrated a 40:1 ratio of coagulated tissue volume to injected material and up to a \SI{30}{\degreeCelsius} temperature increase with thermoembolization \cite{fuentes2020mathematical}. For such a treatment strategy to successfully translate to the clinic, it must be optimized by characterizing and understanding the delivery of the therapy under varying protocols with a range of thermal and chemical stressors.

Mathematical modeling of thermoembolization provides the means to simulate multiple treatment environments. Modeling this treatment modality involves accounting for complex chemically reacting multicomponent flows within porous living tissue. Because thermoembolization can lead to vascular destruction and nonperfused tissue in vivo, it is governed by nonlinear, coupled, and degenerate equations. Mixture theory formulations provide a framework for developing a unified model of chemically reacting mass transport within vascularized, porous living tissue. Various fields, including petroleum engineering and geosciences, have contributed to this approach \cite{peaceman1977fundamentals, faust1979geothermal, bai2003dynamics, soltani2011numerical, khaled2003role, salama2011viscous, tapani1996effect, boucher1998intratumoral, magdoom2014mri, barauskas2007finite, lima2017selection, cristini2009nonlinear, cristini2010multiscale, hawkins2012numerical, lima2014hybrid, lima2015analysis, hawkins2013bayesian,  hawkins2011toward, oden2013selection, oden2015toward, oden2010general, anderson2006hypersonic, kee2005chemically}. Our initial thermoembolization model, which we described previously \cite{fuentes2020mathematical}, couples continuum-scale porous media models with chemically reacting flow. Experiments using an ex vivo kidney demonstrated the crucial role of vascular geometry in model predictions. However, translating this model to \emph{in-vivo} thermoembolization presents challenges owing to the lack of natural blood flow in \emph{ex-vivo} samples.

Given that chemical reactions originating from blood vessels drive the process and considering that vessels are smaller than organs, 1D models of vasculature have proven cost-effective and sufficiently accurate when coupled with 3D tissue models for modeling and simulations \cite{acosta2017cardiovascular, puelz2017comparison}. This approach offers computational benefits while maintaining bounded modeling errors \cite{masri2023modelling, laurino2019derivation}. However, whereas coupled 1D and 3D models offer promising avenues for future research, in the present study, we focused solely on 1D modeling of thermoembolization in a segmented hepatic artery. This approach allows for investigation of the fundamental dynamics of the treatment within the vascular network while setting the stage for use of more complex coupled models in future work. Herein we emphasize data acquisition, image segmentation, and development of a 1D model that captures the essential features of thermoembolization in the hepatic vasculature. 

\section{Methods}
\subsection{Animal Protocol}

This study was conducted at the University of Texas MD Anderson Cancer Center, Houston, TX, USA under an institutionally approved protocol (IACUC protocol number 00001478-RN03 Approved 8/20/2024 by MD Anderson Cancer Center Institutional Animal Care and Use Committee) using three outbred swines. The animals were acclimated and housed according to institutional policy. After induction and intubation, anesthesia was maintained with \SI{2}{\percent} isoflurane, and supplemental oxygen was provided as needed. Buprenorphine was administered at \SI{0.02}{\milli\gram\per\kilogram} intramuscularly for analgesia. Following each experimental procedure, animals recovered and were monitored until return to baseline activity and food intake. Euthanasia after completion of the study was performed via overdose of phenytoin and pentobarbital given intravenously while animals were under general anesthesia. An iodinated contrast medium (Visipaque 320; GE Healthcare, Milwaukee, WI) was used directly as a supplied as a contrast agent for CT scans.

\subsection{Image Acquisition}
Images of the animals were acquired using a 128-slice computed tomography (CT) system (SOMATOM Definition Edge; Siemens Healthineers, Forchheim, Germany). The CT scanner was part of a hybrid suite used in combination with an Artis Q angiography unit (Siemens Healthineers). Pretreatment/postcontrast and posttreatment/precontrast CT images were obtained. CT hepatic arteriography was performed by inserting a microcatheter into the common hepatic artery, injecting the contrast medium, and scanning after a suitable delay. All scans were acquired with a tube voltage of 120 kVp, rotation time of \SI{0.5}{\second}, pitch of 0.6, and 350 effective \SI{}{\milli\ampere\second}. The resulting volumetric CT dose index for each scan was \SI{23.45}{\milli\gray}. Reconstructions were performed at a slice thickness and interval of \SI{0.5}{\milli\meter}, display field of view of \SI{420}{\milli\meter}, and corresponding in-plane pixel size of \SI{0.8}{\milli\meter}.

\begin{figure}[!b]

	\begin{center}
		\includegraphics[width=3.5in]{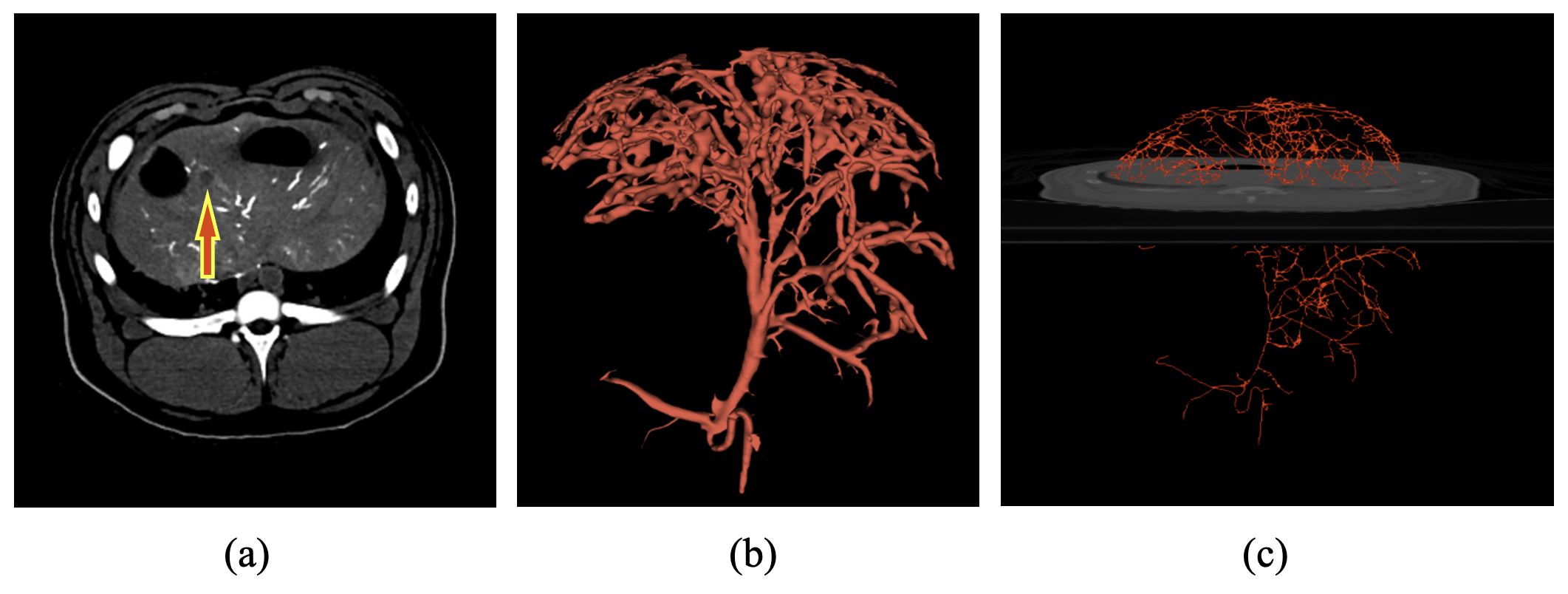}\\
		\caption{Protocol differences and timing of the CT acquisition change the enhancement of arterial vasculature and can impact image segmentation. (a) The arterial phase of a CTHA in a pig model is shown. The red arrow indicates the location of the target lesion. Our imaging protocols provide good contrast between the blood vessels and background liver. (b) Blood vessels are segmented using a Hessian-based vesselness filter. (c) A 3D model of the vasculature 1D centerlines is displayed with respect to the image.}
		\label{fig:imgSeg}
	\end{center}
\end{figure}

\subsection{Image Processing}
Vessel segmentation was performed with a Hessian-based vesselness filter \cite{frangi1998multiscale}. The parameter value thresholds for the blob structure, plate structure, and second-order structures were 0.5, 0.5, and 5.0, respectively, with a Gaussian blur radius of \SI{2}{\milli\meter}. The vesselness filter was set at a threshold of 1 to generate the corresponding segmentation images. A 3D thinning algorithm was applied to extract the vessel centerline \cite{homann2007implementation}. A sign distance transform was applied to computation of the radial distance from the segmentation boundary and the vessel centerline. Landmarks were manually placed on vessel bifurcations of pretreatment/postcontrast imaging and posttreatment/precontrast images. Depending on the visible anatomy in the images, 5-10 landmarks were placed on each image. Landmark-based registration was applied to alignment of the pretreatment and posttreatment images. Fig.~\ref{fig:imgSeg} shows the centerline extraction of the hepatic artery from the arterial phase of CTHA.

\subsection{Converting an NifTi File to a 1D Vascular Network}
The centerline extracted from the segmentation of the arterial phase of CTHA is stored in NifTi file format. The NIfTI file contains centerline voxels, information about the vessel radius, and ethiodized oil (Lipiodol) distribution in the vasculature. A Python script processes the NIfTI file to identify these centerline voxels. The script then employs a connectivity algorithm to construct the vascular tree as follows:
1.	The Python script examines each voxel's immediate neighborhood, considering all adjacent voxels (including diagonals) within a one-voxel distance.
2.	Connected voxels are linked, and the length of each connection is calculated.
3.	This process continues iteratively until all possible voxel connections are established.
Because of limitations in imaging resolution or data quality, the result of this script may not be a single continuous structure. Instead, multiple disconnected trees may form with gaps in the data that prevent direct connections. In such cases, the largest continuous tree (i.e., the one with the most connected voxels) is selected for further analysis provided it accurately represents the hepatic artery. This approach ensures that the simulations are based on the most complete and representative vascular structure available from each imaging dataset while maintaining the integrity of the analysis by excluding inadequate reconstructions.

\begin{figure}[!h]
	
	\begin{center}
		\includegraphics[width=3.5in]{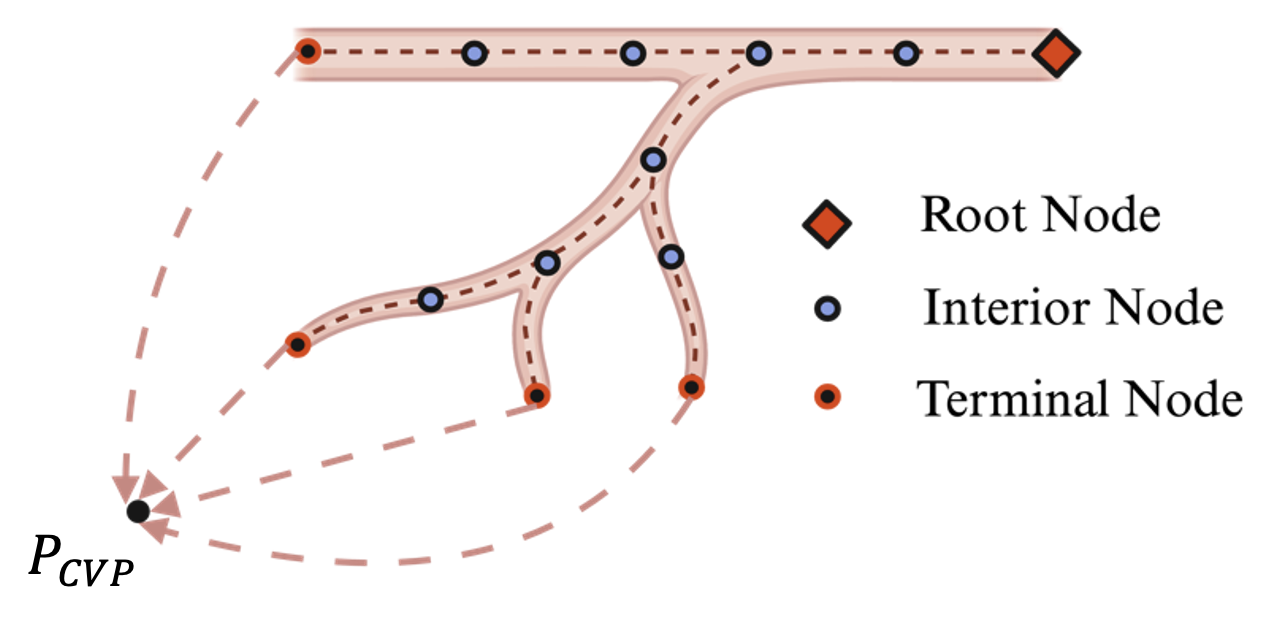}\\
		\caption{Illustration of the root nodes, interior nodes and terminal nodes. The dotted arrows represents the unsegmented virtual blood vessels extending to the sink boundary condition. Central venous pressure was used as the sink pressure for this study.}
		\label{fig:mathModel}
	\end{center}
\end{figure}

\subsection{Mathematical Modeling}
The Hagen-Poiseuille equation (Eq.~\eqref{eq:HagenPoiseuille}) was used to model the 1D blood flow in the segmented vasculature. Each pixel of the centerline was used as a pressure node, and the connection between two pixels is a branch segment. In the segmented vasculature, the pressure nodes can be classified into three categories: interior nodes, terminal nodes, and root nodes. Fig.~\ref{fig:mathModel} provides an illustration of the root nodes, interior nodes, and terminal nodes.

Each node `i' is connected to a set of neighboring nodes $N_i$. Terminal nodes and root nodes are the ones nodes where there is only one neighboring node is connected to `i’. Root nodes are the pressurized terminal nodes where the Dirichlet boundary condition of inlet pressure is imposed as shown in Eq.~\eqref{eq:inlet}. The Neumann boundary condition for the flow continuity is imposed on the terminal nodes as shown in Eq.~\eqref{eq:outlet}. Interior nodes have more than one neighbor, and a mass balance equation is imposed on all of them as shown in Eq.~\eqref{eq:massBalance}.

\begin{eqnarray}
	\label{eq:HagenPoiseuille}
	q_{ij} = k_{ij}(P_i - P_j)
\end{eqnarray}

where,

\begin{eqnarray}
	k_{ij} = \frac{\pi R^4_{ij}}{8 \mu L_{ij}} 
\end{eqnarray}

\begin{eqnarray}
	\label{eq:massBalance}
	\sum_{i \in N_j} (k_{ij}(P_i - P_j)) = 0
\end{eqnarray}

\begin{eqnarray}
	\label{eq:inlet}
	P_{root} = P_{MAP}
\end{eqnarray}

\begin{eqnarray}
	\label{eq:outlet}
	\sum_{j\in N_i} (k_{ij}(P_i - P_j)) + \frac{\gamma_a}{\mu} (P_{CVP} - P_i) = 0
\end{eqnarray}

\begin{table}[!b]
	\caption{Simulation parameters for 1D thermoembolization modeling}
	\label{tab:simPara}
	\centering
	\small
	\begin{tabular}{@{}lcc@{}}
		\toprule
		Parameter & Symbol & Value \\
		\midrule
		Mean arterial pressure & $P_{MAP}$ & \SI{100}{\milli\meter Hg} \\
		Central venous pressure & $P_{CVP}$ & \SI{5}{\milli\meter Hg} \\
		Blood density & $\rho_b$ & \SI{1045}{\kilo\gram\per\cubic\meter} \\
		Blood viscosity & $\mu_b$ & \SI{8.9E-4}{\pascal\second} \\
		Blood specific heat & $c_{p,b}$ & \SI{3600}{\joule\per\kilogram\per\kelvin} \\
		Bolus density & $\rho_o$ & \SI{1280}{\kilo\gram\per\cubic\meter} \\
		Bolus viscosity & $\mu_o$ & \SI{7E-4}{\pascal\second} \\
		Bolus specific heat & $c_{p,o}$ & \SI{1970}{\joule\per\kilogram\per\kelvin} \\
		DCACl density & $\rho_{DCACl}$ & \SI{1532}{\kilogram\per\cubic\meter} \\
		Molarity of DCACl & $M$ & \SI{2}{M} \\
		Saturation of DCACl in bolus & $\epsilon$ & \num{0.1919} \\
		Exothermic energy release & $h$ & \SI{138}{\kilo\joule\per\mole} \\
		\bottomrule
	\end{tabular}
\end{table}

In Eq.~\eqref{eq:outlet}, the pressure drop parameter $\gamma_a$ represents the overall flow conductivity of the unsegmented vasculature between the terminal nodes and the sink term. In this model, the central venous pressure $(P_CVP)$ is used as the sink pressure. The value of $\gamma_a$ plays an important role in ensuring that the total liver perfusion is within the acceptable range. The clinical data regarding the actual liver perfusion in each pig in our study was unknown, but the total pig weights were known. As reported regarding porcine hepatic perfusion, the mean ($\pm$ SD) pig liver weight is about \SI{2.04}{\percent} $\pm$ \SI{0.33}{\percent} of the total weight, and the mean regional blood flow is \SI{22.28}{\milli\liter\per\minute\per 100 \gram} of tissue weight \cite{lin2020physiological}. Using these correlations, the liver weight in each pig and the acceptable range of perfusion were calculated and are shown in Table~\ref{tab:pigDetails}.

\begin{table}[!h]
	\caption{Study pig details}
	\label{tab:pigDetails}
	\centering
	\small
	\begin{tabular}{@{}lcccc@{}}
		\toprule
		\makecell{Pig\\number} & \makecell{Weight \\ (\SI{}{\kilogram})} & \makecell{Min $q_{ref}$ \\ \SI{}{\milli\liter\per\minute}} & \makecell{Max $q_{ref}$ \\ \SI{}{\milli\liter\per\minute}}  & \makecell{Injected bolus \\ (\SI{}{\micro\liter})} \\
		\midrule
		1 & 54 & 205.73 & 285.14 & 200 \\
		2 & 29 & 110.48 & 153.13 & 250 \\
		3 & 34 & 129.53 & 179.53 & 400 \\
		\bottomrule
	\end{tabular}
\end{table}

The optimal value of $\gamma_a$ was determined using the inlet boundary condition as the mean arterial pressure and the outlet pressure as the central venous pressure. Eq.~\eqref{eq:optimizeGa} is the objective function used to minimize the error between the reference liver perfusion and simulated liver perfusion (Eq.~\eqref{eq:minF}). The Nelder-Mead method of optimization was used to find the optimalum value of $\gamma_a$ with a tolerance value of \num{1E-16}. The \emph{scipy.optimize} library of in Python was used to perform the Nelder-Mead optimization. 

\begin{eqnarray}
	\label{eq:minF}
	f(\gamma_a) = || q_{calc}(\gamma_a) - q_{ref}||^2
\end{eqnarray}

\begin{eqnarray}
	\label{eq:optimizeGa}
	\min_{\gamma_a} f(\gamma_a)
\end{eqnarray}

The saturation of the bolus ($s_o$); (consisting of Lipiodol and dichloroacetyl chloride (DCACl)) was modeled in a segmented \emph{in -vivo} hepatic artery using a 1D advection equation as shown in Eq.~\eqref{eq:massTransport}. The right-hand side of Eq.~\eqref{eq:massTransport} represents the chemical reaction that happens in the vasculature, which results in DCACl undergoing hydrolysis, producing additional acids and heat. For simplicity, the Lipiodol and DCACl are not modeled separately, and the entire bolus is considered as homogonous mixture undergoing the hydrolysis. The limitations of this assumptions are described in the Discussion section. In Eq.~\eqref{eq:massTransport}, the $\epsilon$ represents the saturation of DCACl in the bolus. In all three pigs, a \SI{2}{M} solution of DCACl was mixed with Lipiodol. Hence, a total saturation of $\epsilon = $ \num{0.1919} was used for all the pigs. However, the amount of bolus injected in each pig varied, and is given in Table \ref{tab:pigDetails}.

\begin{eqnarray}
	\label{eq:massTransport}
	\frac{\partial s_o}{\partial t} + \frac{\partial u s_o}{\partial x} = - \frac{\gamma_t \epsilon \rho_{DCACl}}{\rho_o} s_o
\end{eqnarray}

\begin{eqnarray}
	\label{eq:temperature}
	\frac{\partial T}{\partial t} + \frac{\partial u T}{\partial x} = h \frac{\epsilon \gamma_t \rho_{DCACl}}{W \rho c_p} s_o
\end{eqnarray}

In Eq.~\eqref{eq:massTransport}, the rate of hydrolysis of DCACl is represented using $\gamma_t$ (\SI{}{\per\second}). This term was determined using curve fitting to temperature values in a previous work by Fuentes et al. [11]. Determining the value $\gamma_t$ and its influence on mass transport is the focus of present study. Eq.~\eqref{eq:temperature} is the energy equation used to calculate the temperature rise owing to the exothermic chemical reaction of DCACl. For this purpose, the saturation $s_o$ of bolus is tracked at every node. By tracking $s_o$, the amount of chemical reaction each node of the vasculature experiences is tracked. The understanding is that the nodes that experience more chemical reaction than others will undergo thermoembolization. The time constant of hydrolysis $\gamma_t$ controls the rate at which this hydrolysis happens. If the hydrolysis happens too quickly, all of the chemical reaction takes place at the injection site. If the time constant is too small, a substantial amount of DCACl escapes the segmented vasculature and will cause damage to the tissue and vasculature further downstream. Hence, one of the criteria for determining $\gamma_t$ is that the amount of bolus escaping without undergoing hydrolysis should be zero. Eq.~\eqref{eq:massTransport} and Eq.~\eqref{eq:temperature} are discretized using the backward Euler upwind scheme and solved. The simulation ends when no more bolus is left in the bloodstream to undergo hydrolysis. Thus, time-stepping stops when no more acid chloride is left to undergo reaction. At every time-step, the amount of hydrolysis reaction experienced by each node of the segmented vasculature is tracked. This is shown in Eq.~\eqref{eq:Dx}, where $V_{DCACl}$ represents the total amount of bolus injected into a pig, $\Delta V_x$ represents the volume of vascular element at x, and $D_x$ represents the cumulative reaction experienced by node x over time period t.

\begin{eqnarray}
	\label{eq:Dx}
	D_x = \frac{1}{V_{DCACl}} \frac{\gamma_t \epsilon \rho_{DCACl}}{\rho_o} \sum_t s_o (x,t) \Delta t \Delta V_x
\end{eqnarray}

\begin{eqnarray}
	\label{eq:damage}
	\mathcal{X}_x = \{x:D_x \geq \delta\}
\end{eqnarray}

All of the vascular segments having $D_x \geq \delta$ are tagged as possible sites of embolization $(\mathcal{X}_x)$. This threshold value of minimum chemical reaction at a given location $(\delta)$ is belived to significantly influence the model's overall predictive accuracy. Two critical parameters are expected to play pivotal roles in our model's predictive capabilities: hepatic arterial blood flow $q_{ref}$ and threshold $\delta$. Given that precise values for these parameters are not definitively known, a comprehensive uncertainty study is conducted. This analysis aims to quantify the impact of these parameters on the model's performance and refine our understanding of their optimal ranges. This approach allows us to account for variability in these key factors and enhance the robustness of our predictive model for possible embolization sites.

\begin{eqnarray}
	\label{eq:opttC}
	g(\gamma_t, \delta) = \beta(\gamma_t) + (1 - \alpha(\gamma_t, \delta))
\end{eqnarray}

\begin{eqnarray}
	\label{eq:mintC}
	\min_{\gamma_t} g(\gamma_t, \delta)
\end{eqnarray}
To determine the optimal value of $\gamma_t$, Eq.~\eqref{eq:opttC} is minimized (Eq.~\eqref{eq:mintC}) using the Nelder-Mead method, where $\beta(\gamma_t)$ represents the percentage of bolus escaping the segmented vasculature without undergoing hydrolysis and $\alpha(\gamma_t, \delta)$ represents the balanced accuracy calculated for the predicted embolization site versus the observed in vivo Lipiodol concentration. To ensure that the final result of optimization (Eq.~\eqref{eq:mintC}) is not affected by any possible local optimas dependent on the initial guess, 10 initial guess values are randomly generated between $\gamma_{t0} \in (10, 20) $. In  Eq.~\eqref{eq:opttC}, the percentage of bolus escaping the segmented vasculature ($\beta$) is a function of $\gamma_t$ alone for a given $q_{ref}$, as the rate of hydrolysis for a given rate of blood flow determines how far the DCACl will flow without undergoing hydrolysis. Furthermore, in comparison the balanced accuracy $\alpha$ depends on the minimum threshold ($\delta$) of the possible embolization site.

\subsection{Uncertainty Analysis}

Uncertainty analysis was performed by varying the hepatic arterial blood flow within the acceptable range for each pig, and the threshold value $\delta$ varied from \SI{1}{\percent} to \SI{10}{\percent}. The minimum and maximum liver perfusion for each pig as per the correlation between its weight and liver perfusion given by Lin et al. [44] are shown in Table \ref{tab:pigDetails}. Ten values of liver perfusion within the acceptable range for each pig and ten values of $\delta \in (\SI{1}{\percent}, \SI{10}{\percent})$ are used for uncertainty analysis. For each value of liver perfusion, the value of $\gamma_a$ is calculated using Eq.~\eqref{eq:minF} and Eq.~\eqref{eq:optimizeGa} such that the simulated total blood flow rate is the same as the expected reference blood flow rate. The optimalum time constant of hydrolysis ($\gamma_t$) is determined for each combination of $q_{ref}$ and $\delta$ values using Eq.~\eqref{eq:opttC} and Eq.~\eqref{eq:mintC}. 

\begin{figure}[!b]
	
	\begin{center}
		\includegraphics[width=3.5in]{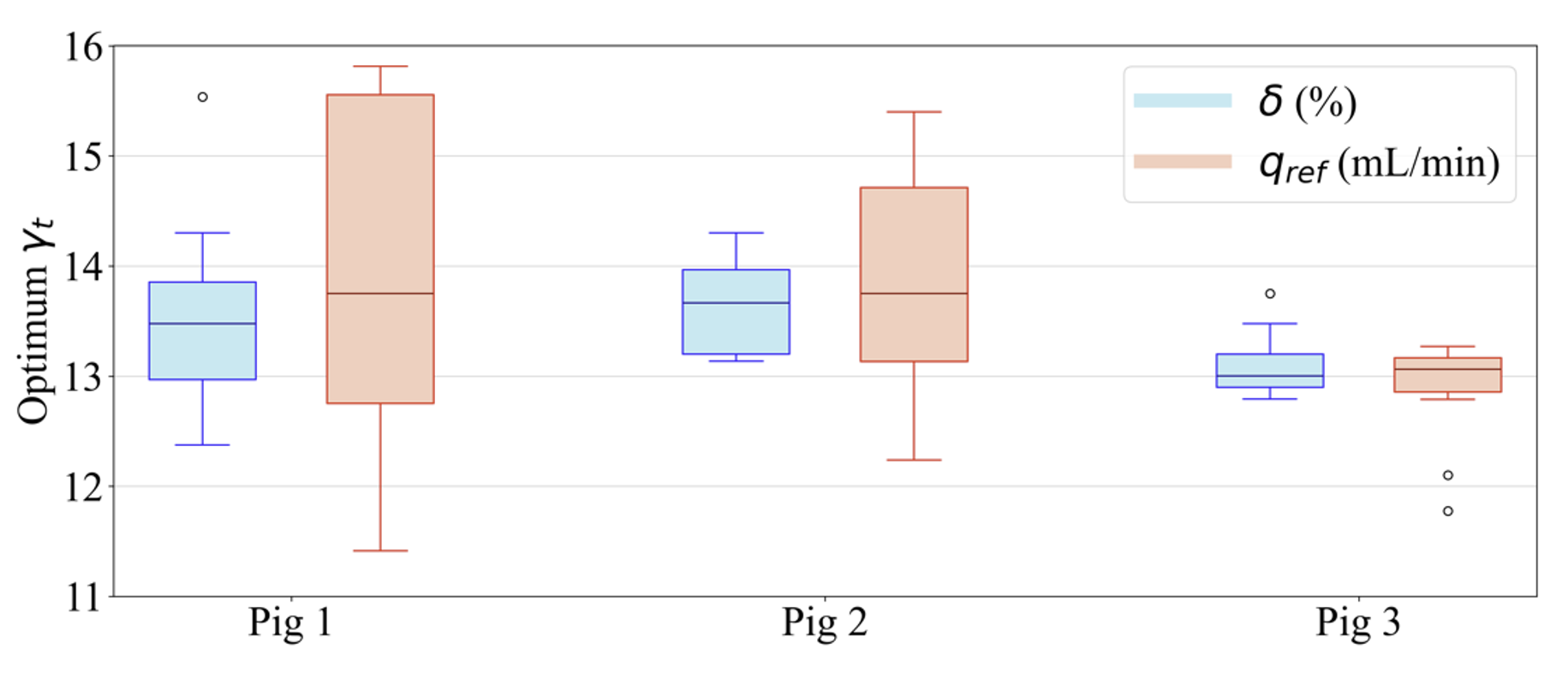}\\
		\caption{Uncertainty analysis of the blood flow rate in the hepatic artery in each pig and the minimum threshold of experienced hydrolysis reaction at each location.}
		\label{fig:uncertAnalysis}
	\end{center}
\end{figure}

\section{Results}

\begin{figure*}[!htb]
	\centering
	\includegraphics[width=0.9\textwidth]{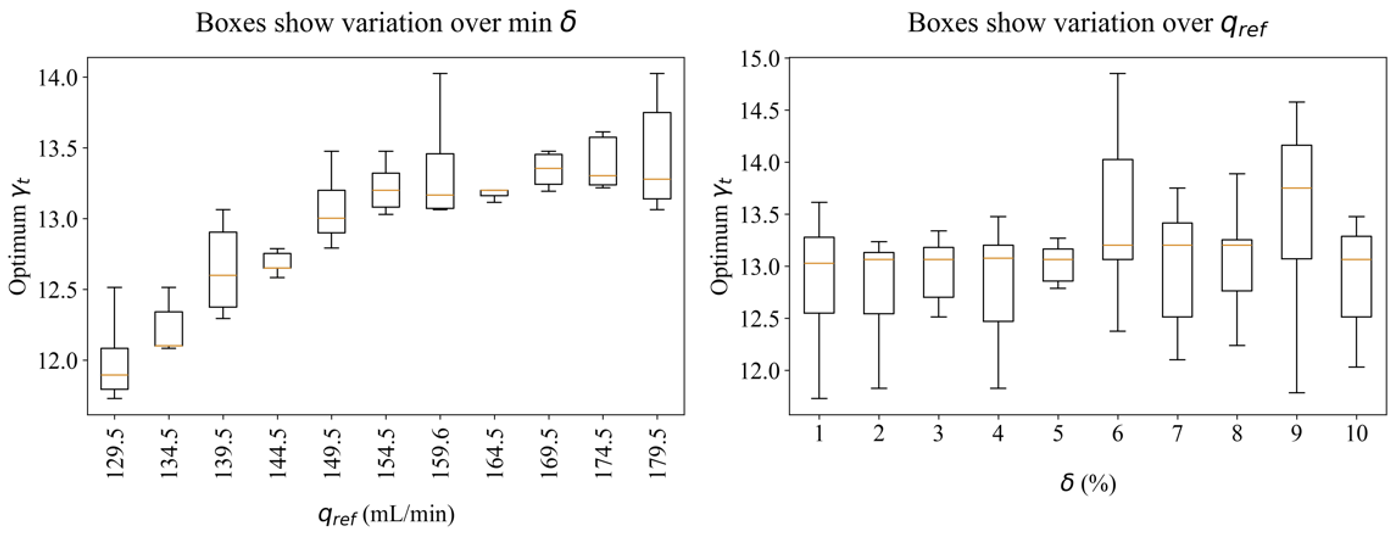}
	\caption{Uncertainty analysis of hepatic arterial blood flow rate (left) and minimum threshold (right) for possible embolization sites for pig 3.}
	\label{fig:A3}
\end{figure*}

A summarized plot of uncertainty analysis results across all three pigs is shown in Fig.~\ref{fig:uncertAnalysis}. More detailed plots for each pig are given in the Appendix (Fig.~\ref{fig:A1}, Fig.~\ref{fig:A2}, and Fig.~\ref{fig:A3}), and the balanced accuracy for the mean of each pig is given in Fig.~\ref{fig:A4}, Fig.~\ref{fig:A5}, and Fig.~\ref{fig:A6} in the Appendix.

The Pearson correlation test was conducted to identify any correlation between $q_{ref}$ and optimal $\gamma_t$ and between the threshold $\delta$ value and optimal $\gamma_t$. The results of this analysis are given in Table~\ref{tab:corr}.

\begin{table}[!t]
	\caption{Pearson correlation analysis results}
	\label{tab:corr}
	\centering
	\small
	\begin{tabular}{@{}lccc@{}}
		\toprule
		& \multicolumn{3}{c}{Optimal $\gamma_t$} \\
		\cmidrule(l){2-4}
		 Variables & Pig 1 &Pig 2 &Pig 3   \\
		\midrule
		$q_{ref}$ & 0.66 & 0.88 & 0.76  \\
		$\delta$ & 0.02 & -0.12 & 0.20  \\
		\bottomrule
	\end{tabular}
\end{table}

The liver perfusion value ($q_{ref}$) appeared to have a consistently strong positive correlation with the optimal $\gamma_t$ across all three pigs. This suggested that $q_{ref}$ is a good predictor of optimal $\gamma_t$ and must be considered in any mathematical equation that may be derived in future work for determining the optimal $\gamma_t$. Higher $q_{ref}$ values generally corresponded to higher optimal $\gamma_t$ values. This intuitively makes sense because if the blood flow rate is high and hydrolysis rate is low, more embolization and possible damage are expected farther downstream.

\begin{figure}[!ht]
	
	\begin{center}
		\includegraphics[width=3.5in]{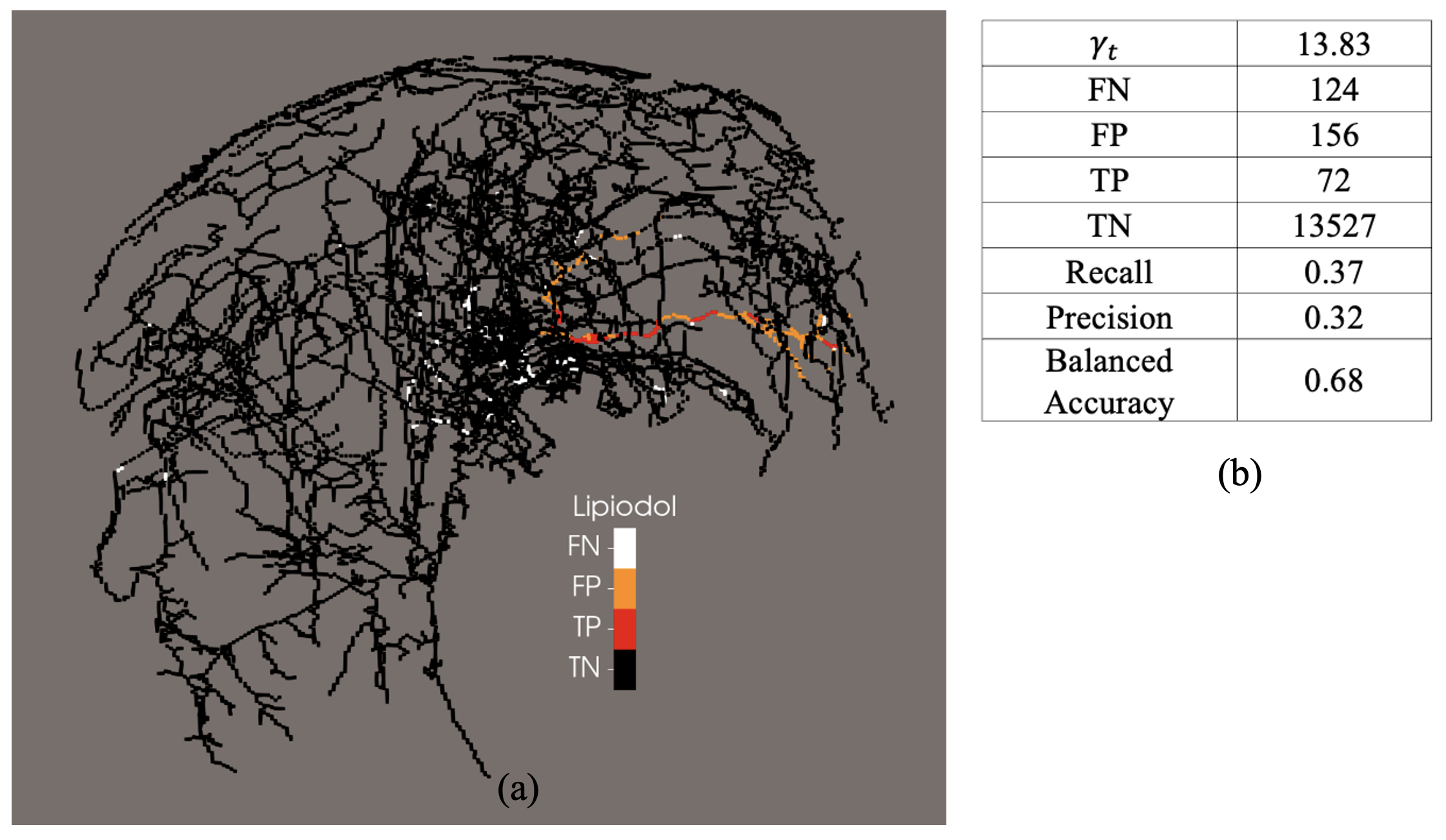}\\
		\caption{{Analysis of the possible embolization site prediction of the 1D model for pig 1}. (a) Predictive ability of the model compared with the in vivo data. (b) Table of the confusion matrix for of the predicted embolization sites and \emph{in vivo} Lipiodol post treatment. FN, false-negative; FP, false-positive; TP, true-positive; TN, true-negative.}
		\label{fig:pig1}
	\end{center}
\end{figure}

\begin{figure}[!ht]
	
	\begin{center}
		\includegraphics[width=3.5in]{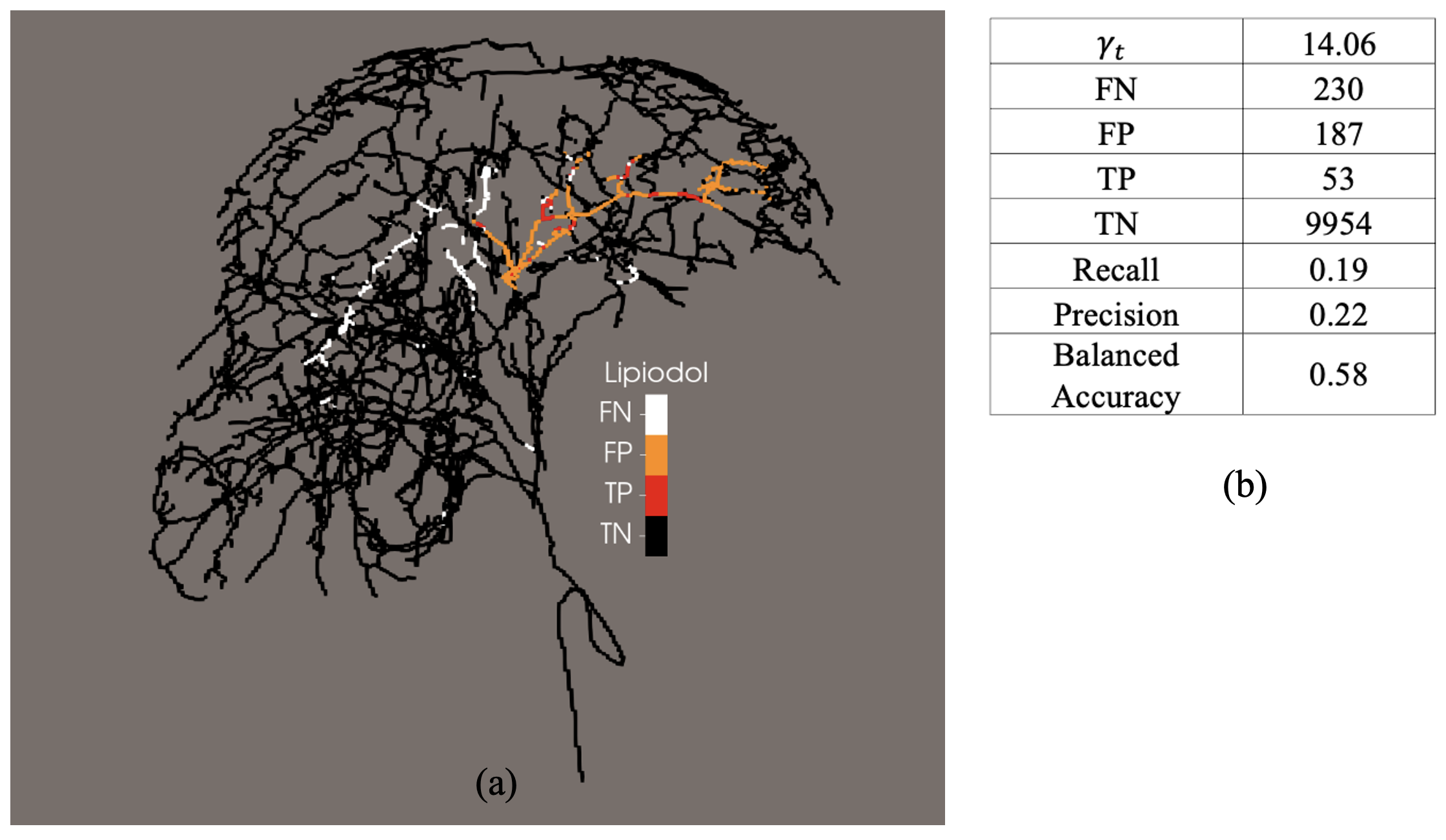}\\
		\caption{{Analysis of the possible embolization site prediction of the 1D model for pig 2}. (a) Predictive ability of the model compared with the in vivo data. (b) Table of the confusion matrix for of the predicted embolization sites and \emph{in vivo} Lipiodol post treatment. FN, false-negative; FP, false-positive; TP, true-positive; TN, true-negative.}
		\label{fig:pig2}
	\end{center}
\end{figure}

\begin{figure}[!ht]
	
	\begin{center}
		\includegraphics[width=3.5in]{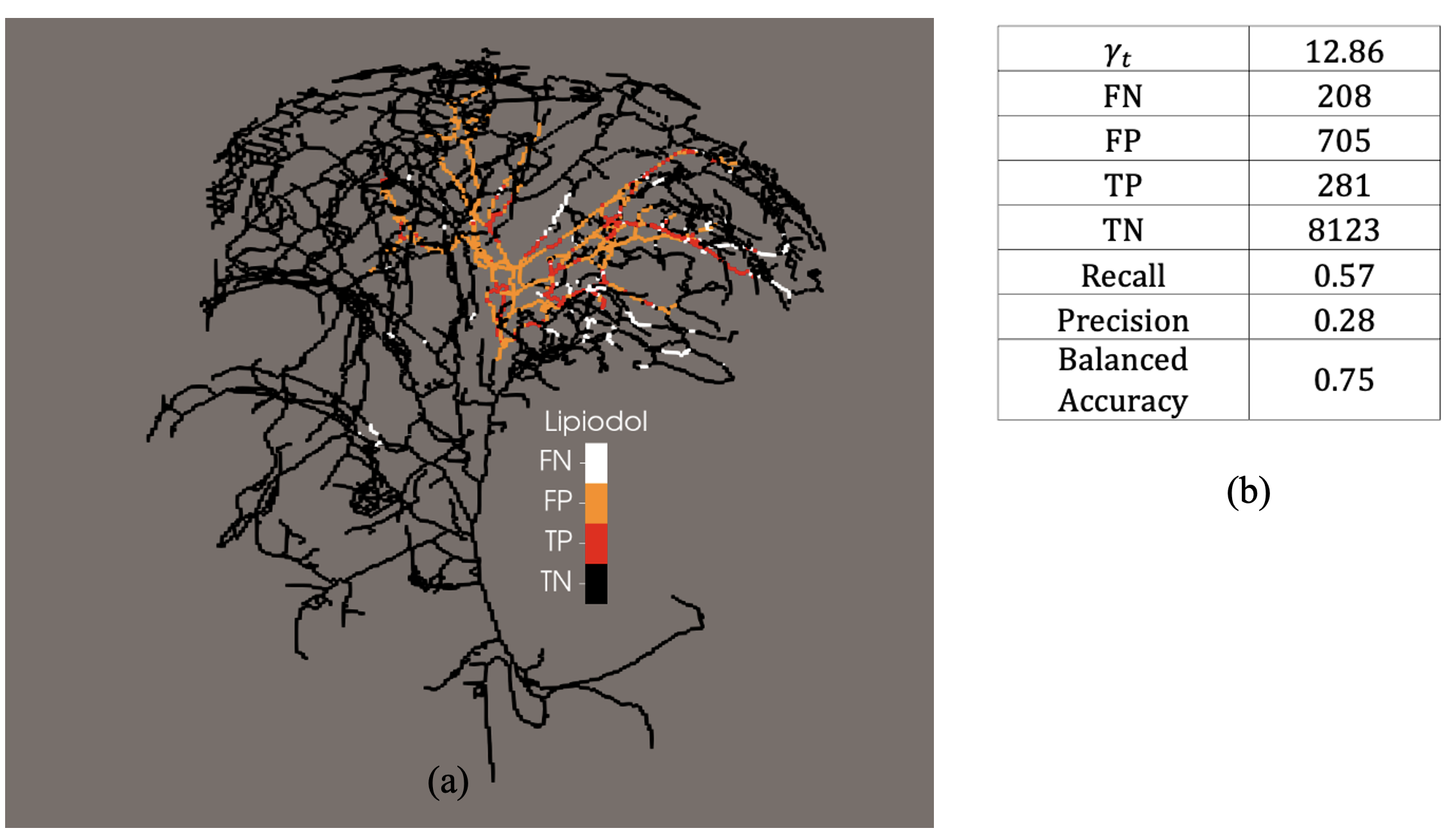}\\
		\caption{{Analysis of the possible embolization site prediction of the 1D model for pig 3}. (a) Predictive ability of the model compared with the in vivo data. (b) Table of the confusion matrix for of the predicted embolization sites and \emph{in vivo} Lipiodol post treatment. FN, false-negative; FP, false-positive; TP, true-positive; TN, true-negative.}
		\label{fig:pig3}
	\end{center}
\end{figure}

\begin{figure*}[!htb]
	\centering
	\includegraphics[width=\textwidth]{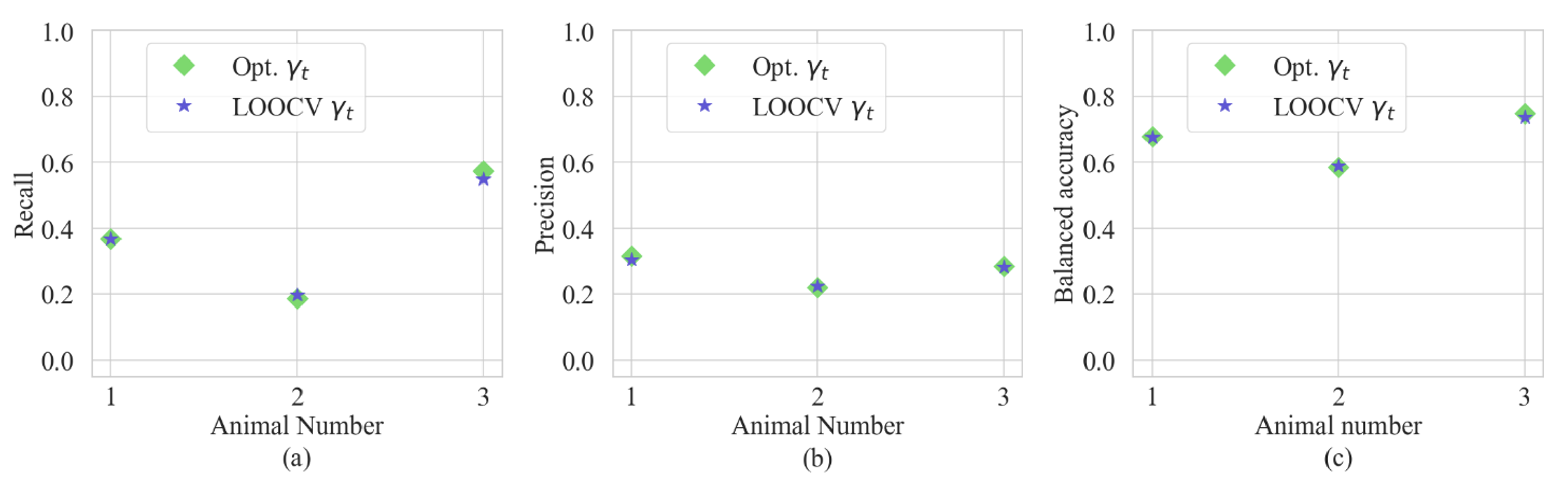}
	\caption{Recall, precision, and balanced accuracy for the optimal calculated $\gamma_t$ and LOOCV $\gamma_t$.}
	\label{fig:recall}
\end{figure*}

In contrast with the correlation between $q_{ref}$ and the optimal $\gamma_t$, we observed an inconsistent and weak correlation between the threshold value $\delta$ and optimal $\gamma_t$. Determining the reason of this correlation requires further analysis, which was beyond the scope of the present study. The correlation between $q_{ref}$ and the optimal $\gamma_t$, and that between $\delta$ and the optimal $\gamma_t$ are depicted visually in Fig.~\ref{fig:A1}, Fig.~\ref{fig:A2}, and Fig.~\ref{fig:A3} for pig 1, pig 2, and pig 3, respectively, in the Appendix.

The observed locations where Lipiodol was stuck \emph{in-vivo} after thermoembolization treatment, and the prediction of our model for the expected Lipiodol clogging and possible embolization sites are shown in Fig.~\ref{fig:pig1}, Fig.~\ref{fig:pig2}, and Fig.~\ref{fig:pig3}, for pig 1, pig 2, and pig 3, respectively.

As shown in Fig.~\ref{fig:pig1}, Fig.~\ref{fig:pig2}, and Fig.~\ref{fig:pig3}, false -negative represents locations in the vasculature at which our model predicted no embolization but we obserbed \emph{in-vivo} embolization, whereas false-positive represents locations in the vasculature at which our model predicted possible embolization but we did not observe in vivo embolization. True-negative and true-positive represents locations in the vasculature at which our model correctly predicted no emoblization and possible embolization, respectively.

As shown in Fig.~\ref{fig:pig2}, we observed more points of the vasculature that were false-negative than pig 1 and pig 3. This is because the segmentation obtained from the imaging data could not reproduce the continuous blood vessel where false-negative findings are dominant. As a result, the path taken by the bolus in our model failed to capture the exact path which it would take \emph{in-vivo}.

\subsection{Cross-validation}
The optimal $\gamma_t$ for each pig 1, pig 2, and pig 3 across different values of $q_{ref}$ and $\delta$ is shown in Fig.~\ref{fig:A1}, Fig.~\ref{fig:A2}, and Fig.~\ref{fig:A3}, respectively, in the Appendix. Similarly, the balanced accuracy $\alpha$ for different values of $q_{ref}$ and $\delta$ are shown in Fig.~\ref{fig:A4}, Fig.~\ref{fig:A5}, and Fig.~\ref{fig:A6}, in Appendix. We observed the maximum balanced accuracy $\alpha$ especially when $\delta = $ \SI{1}{\percent}. Hence, to perform a cross-validation study, we used the mean $q_{ref}$ and used $\delta = $ \SI{1}{\percent} for each pig to find the $\gamma_t (q_{ref}, \delta) $. Specifically, we used the leave-one-out cross-validation (LOOCV) method to cross-validate the model performance across all three pigs. 
Fig.~\ref{fig:recall} shows the recall, precision, and balanced accuracy calculated to quantify the predictive ability of our model when compared with in vivo data. We calculated each of these {“values”? “parameters”?} for the optimal value of $\gamma_t$. This would represent the best possible prediction of {“using”?} our model for each pig. Using LOOCV, we calculated these three values and compared them with the optimal values as shown in Fig.~\ref{fig:recall}. 

When compared in terms of balanced accuracy, the LOOCV values and optimal values did not differ substantially. The average balanced accuracy rate for our model using  LOOCV was \SI{66.8}{\percent} when calculated for all three pigs. The $\gamma_t$ calculated for each pig and the LOOCV $\gamma_t$ values are compared in Table~\ref{tab:tCEachPig}.

\begin{table}[!h]
	\caption{Time constant values ($\gamma_t$) for each pig}
	\label{tab:tCEachPig}
	\centering
	\small
	\begin{tabular}{@{}lccc@{}}
		\toprule
		Variables & Pig 1 &Pig 2 &Pig 3   \\
		\midrule
		Optimal $\gamma_t$ & 13.83 & 14.06 & 12.86  \\
		LOOCV $\gamma_t$ & 13.45 & 13.35 & 13.94  \\
		\bottomrule
	\end{tabular}
\end{table}

\section{Discussion}
Considering the assumptions made to simplify this model for preliminary analysis, an overall balanced accuracy rate of \SI{66.8}{\percent} is very promising.

In our previous work, we modeled thermoembolization using an ex vivo kidney \cite{fuentes2020mathematical} and demonstrated a substantial increase in the temperature of tissue owing to the thermoembolization. In the present study, we simplified that model and extended it to an \emph{in vivo} hepatic artery. We solved the temperature equation (Eq.~\eqref{eq:temperature}) to determine the temperature rise owing to the chemical reaction between DCACl and tissue. However, our model predicted a very modest temperature rise of up to \SI{0.1}{\degreeCelsius} for the hydrolysis reaction in the hepatic artery \emph{in vivo}. This is to be expected after considering the much smaller amount of DCACl-Lipiodol solution delivered in the vessels compared to that in kidney \cite{fuentes2020mathematical}. The small temperature increase predicted by our model can be attributed to the complex biophysics of thermoembolization. Our current understanding regarding in vivo thermoembolization is that the exothermic chemical reaction will matter at the endothelial cell level and a few hundred microns deeper from the vasculature in the surrounding tissue. A dedicated \emph{in vivo} study in which the temperature rise produced by thermoembolization can be tracked would provide additional insight and could validate our temperature results.

\begin{figure}[!h]
	
	\begin{center}
		\includegraphics[width=3.5in]{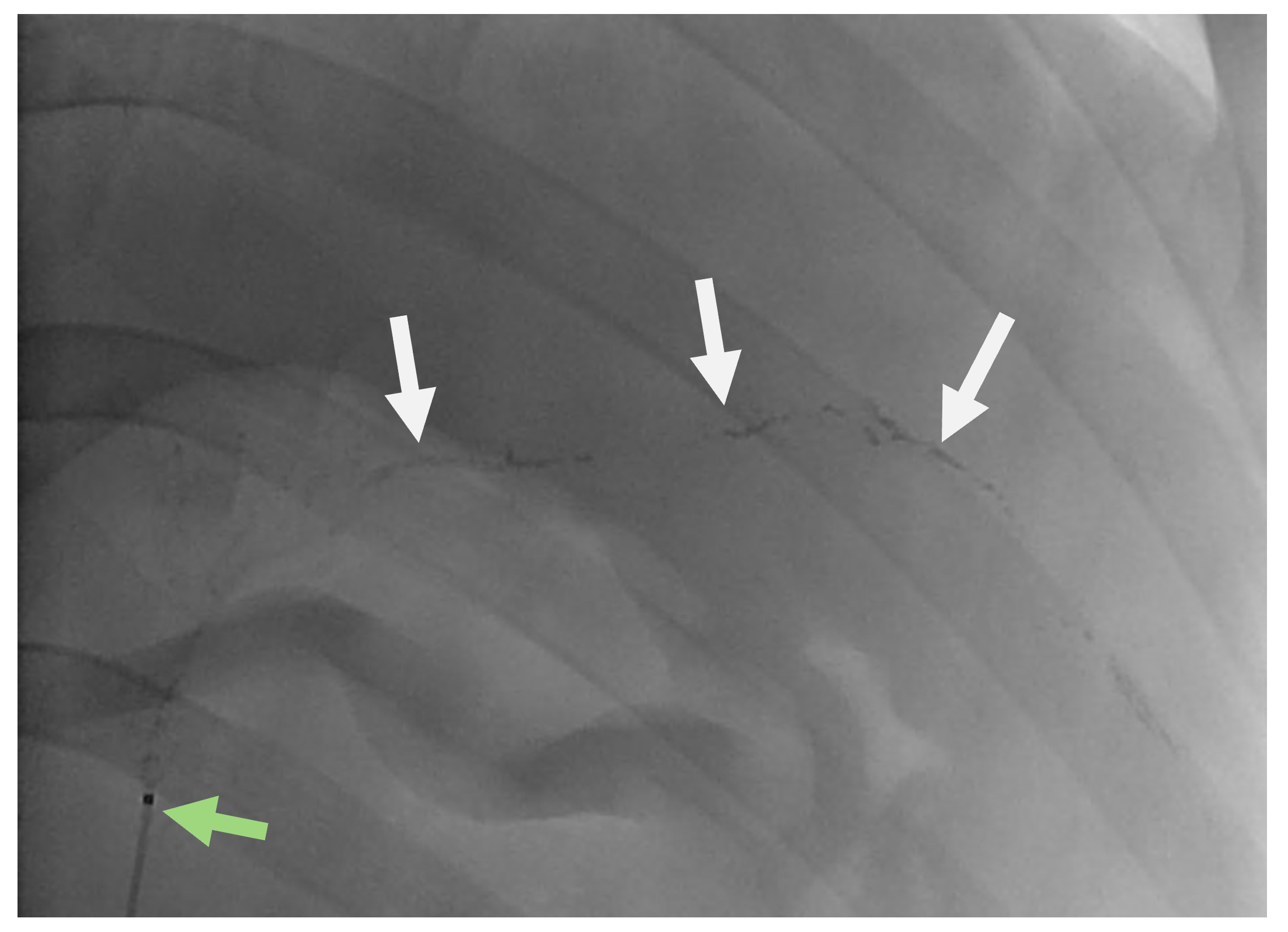}\\
		\caption{Fluoroscopic image during delivery of a DCACl-Lipiodol solution. The globular behavior of the bolus is readily appreciated (white arrows). The tip of the microcatheter is indicated by the green arrow.}
		\label{fig:ctScan}
	\end{center}
\end{figure}

\begin{figure}[!h]
	
	\begin{center}
		\includegraphics[width=3.5in]{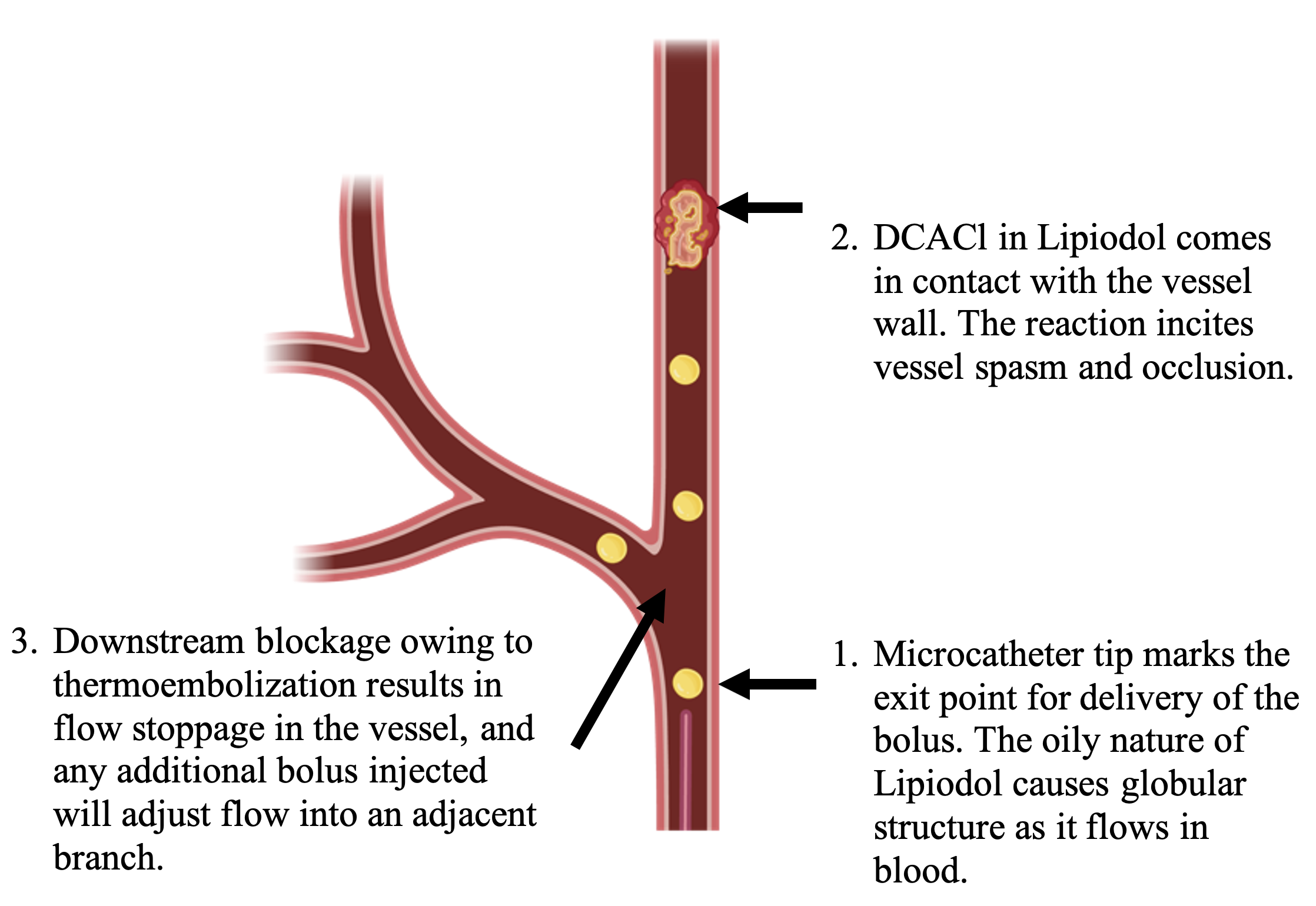}\\
		\caption{The process of thermoembolization observed in the vasculature \emph{in vivo}}
		\label{fig:invivo}
	\end{center}
\end{figure}

Miscible flow approximation about the bolus and blood flow was a limitation of our study. The observed in vivo behavior bolus in oil is immiscible flow. In the blood flow, the bolus of hydrophobic material forms globular immiscible droplets that are carried along by the blood flow as shown by the white arrow in Fig.~\ref{fig:ctScan}. These droplets contain DCACl dissolved in Lipiodol. The chemical hydrolysis reaction \emph{in-vivo} theoretically starts immediately on the surface of the droplets as they come in contact with blood, which is aqueous and contains many potentially reactive proteins. As the droplets enter smaller vessels, they come in contact with the vessel walls. As this happens, the DCACl solution slows and quickly stops forward flow. Stasis then increases the dwell time of the bolus, and under the tested conditions, the effects of thermoembolization appear irreversible. This is illustrated in Fig.~\ref{fig:invivo}.

The 1D flow model we used to predict the possible embolization sites in vessels is preliminary and needs modeling off various complexities to be able to predict the actual biophysics occurring during in vivo thermoembolization. In various studies, modeling of embolization owing to products such as Onyx \cite{orlowski2012computational} and Lipiodol \cite{white2013computational} simulated increases in the viscosity of the bolus as a feedback function. Onyx is a mixture of ethylene vinyl alcohol, dimethyl sulfoxide, and traces of suspended tantalum powder. Once injected, the dimethyl sulfoxide solvent dissipates into the blood and interstitial fluids, causing the ethylene vinyl alcohol copolymer and tantalum to precipitate in situ. This precipitation transforms Onyx from a liquid to a spongy, coherent embolus that solidifies from the outside to the inside, allowing for a slow, controlled injection. In the model described by Orlowski et al. \cite{orlowski2012computational}, an increase in viscosity  of bolus was modeled as an inverse function of the dimethyl sulfoxide concentration. As the dimethyl sulfoxide concentration decreases, the viscosity increases. This simplification helps model the basics of embolization of cerebral arteriovenous malformations using Onyx but does not completely capture the biophysics that result in embolization from Lipiodol with DCACl in the present study. The globular nature of immiscible oil in blood, the delay in the initiation of hydrolysis of DCACl, and the blocking of blood vessels downstream from injection site owing to constriction of blood vessels play a crucial role in determining where the actual damage to tissue and tumor occurs. In addition, the effect of exothermic chemical reactions on embolization must be studied in detail. Furthermore, when compared with existing treatments like transarterial chemoembolization, does the exothermic component of thermoembolization benefit this novel treatment, and to what extent must be analyzed further. Computational models can be excellent tools for studying this, and a high-fidelity complex fluid model is required to capture this biophysics to predict the damaged region with extremely high accuracy.

Moreover, 1D thermoembolization models based on segmented imaging data are highly dependent on the accuracy of the segmentation and the levels of bifurcation of vessels that are segmented. Physics-based models like ours rely on the radius and length information for the segmented blood vessel to predict blood flow patterns and the resultant bolus distribution. Segmentation errors are bound to reduce the accuracy of a model. Because of this, multicompartment porous domains and simulated 3D domains are preferred for many physics-based models. Understanding how such a 3D porous domain can be used to simulate thermoembolization with accurate prediction of damaged regions tissue and tumor will be covered in our future work.

\section{Conclusion}
Herein we present a preliminary mathematical model for predicting the effects of thermoembolization on the hepatic artery in HCC treatment. Our simplified 1D Hagen-Poiseuille blood flow model achieved a promising balanced accuracy rate of \SI{66.8}{\percent} in identifying potential embolization locations. This result is encouraging considering the model's limitations and assumed simplifications. The model's performance suggests that even basic computational approaches can provide valuable insight into the complex biophysics of thermoembolization. However, several key areas require further investigation and refinement.

Whereas our model predicted only a modest temperature rise of \SI{0.1}{\degreeCelsius} for the hydrolysis reaction in the hepatic artery, the localized impact of the exothermic chemical reaction in thermoembolization on endothelial cells and surrounding tissue warrants dedicated in vivo studies. Also, the miscible flow approximation in the current model must be updated to account for the observed immiscible globular nature of the hydrophobic material in blood. A more sophisticated model is needed to capture the delayed initiation of DCACl hydrolysis and its interaction with blood components. 

Despite these limitations, our model provides a foundation for understanding the vascular transport phenomenon in thermoembolization. This work represents a crucial step toward developing a comprehensive computational framework for thermoembolization. Such a framework could significantly enhance treatment planning and optimization for HCC and other diseases amenable to this innovative therapy. By continuing to refine and validate such models, we aim to provide clinicians with powerful tools to make informed decisions about and improve patient outcomes of minimally invasive cancer treatments.

\bibliographystyle{IEEEtran}
\bibliography{IEEEabrv,Bibliography}
%


\onecolumn
\clearpage
\section{Appendix}

\begin{figure*}[!htb]
	\centering
	\includegraphics[width=0.9\textwidth]{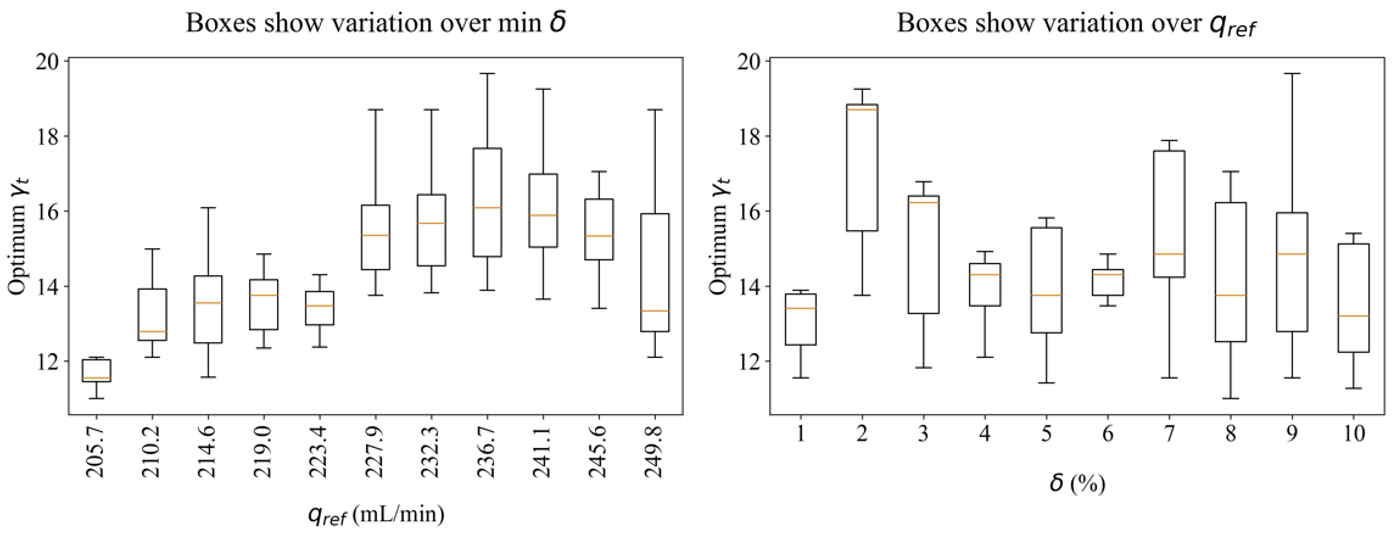}
	\caption{Uncertainty analysis of hepatic arterial blood flow rate (left) and minimum threshold (right) for possible embolization sites for pig 1.}
	\label{fig:A1}
\end{figure*}

\begin{figure*}[!htb]
	\centering
	\includegraphics[width=0.9\textwidth]{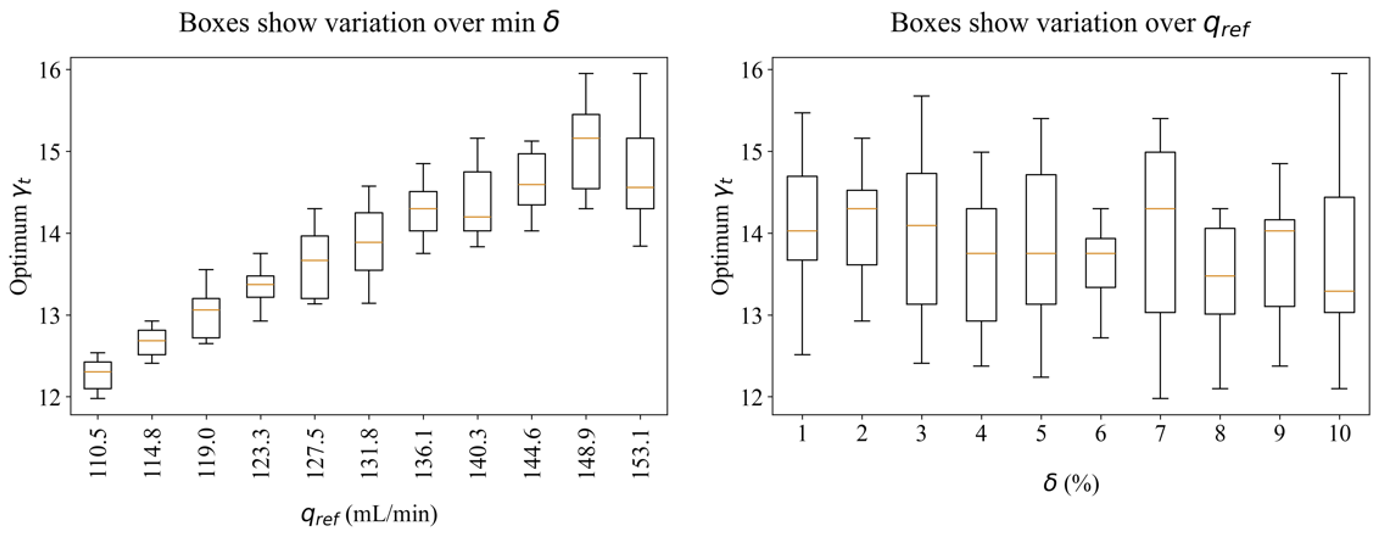}
	\caption{Uncertainty analysis of hepatic arterial blood flow rate (left) and minimum threshold (right) for possible embolization sites for pig 2.}
	\label{fig:A2}
\end{figure*}

\begin{figure*}[!htb]
	\centering
	\includegraphics[width=\textwidth]{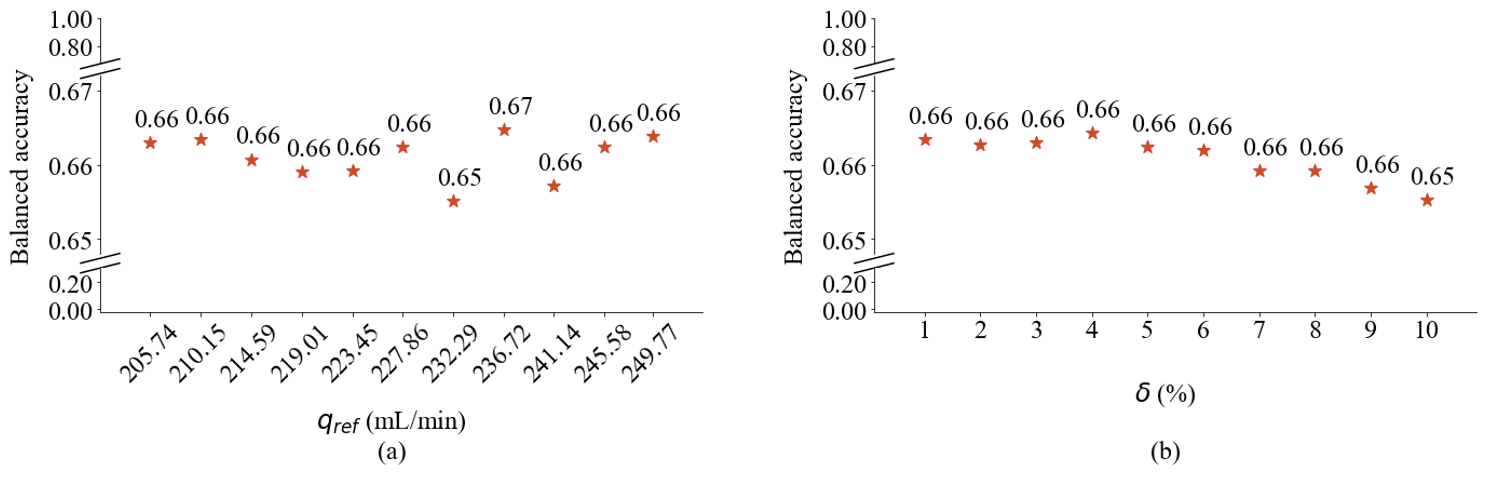}
	\caption{Balanced accuracy analysis for pig 1. (a) Balanced accuracy of the 1D model as a function of variable $q_{ref}$. Each point represents the balanced accuracy calculated using the mean $\gamma_t$ value across all $\delta$ values for a given $q_{ref}$. (b) Balanced accuracy of the 1D model as a function of variable $\delta$. Each point represents the balanced accuracy calculated using the mean $\gamma_t$ value across all $q_{ref}$ values for a given $\delta$.}
	\label{fig:A4}
\end{figure*}

\begin{figure*}[!htb]
	\centering
	\includegraphics[width=\textwidth]{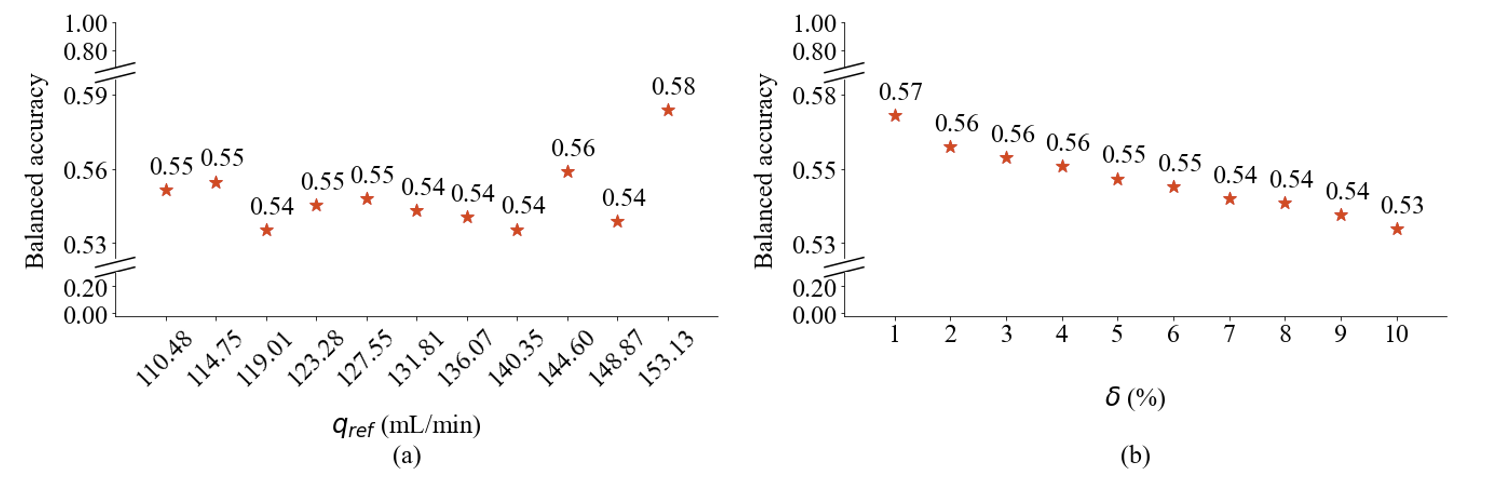}
	\caption{Balanced accuracy analysis for pig 2. (a) Balanced accuracy of the 1D model as a function of variable $q_{ref}$. Each point represents the balanced accuracy calculated using the mean $\gamma_t$ value across all $\delta$ values for a given $q_{ref}$. (b) Balanced accuracy of the 1D model as a function of variable $\delta$. Each point represents the balanced accuracy calculated using the mean $\gamma_t$ value across all $q_{ref}$ values for a given $\delta$.}
	\label{fig:A5}
\end{figure*}

\begin{figure*}[!htb]
	\centering
	\includegraphics[width=\textwidth]{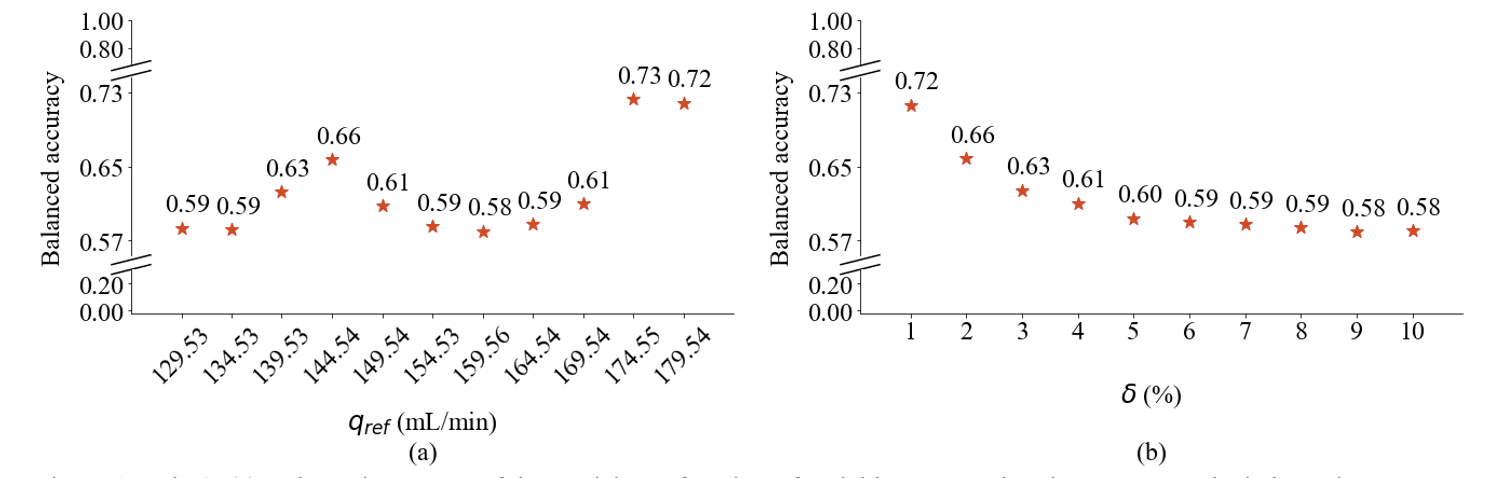}
	\caption{Balanced accuracy analysis for pig 3. (a) Balanced accuracy of the 1D model as a function of variable $q_{ref}$. Each point represents the balanced accuracy calculated using the mean $\gamma_t$ value across all $\delta$ values for a given $q_{ref}$. (b) Balanced accuracy of the 1D model as a function of variable $\delta$. Each point represents the balanced accuracy calculated using the mean $\gamma_t$ value across all $q_{ref}$ values for a given $\delta$.}
	\label{fig:A6}
\end{figure*}

\end{document}